\documentclass[floats, amsmath,nofootinbib, superscriptaddress, preprintnumbers, prd]{revtex4}
\usepackage{color,hyperref}
\usepackage[active]{srcltx}
\usepackage{amsmath,amsfonts,amssymb,amsthm,amstext,amscd,eucal,srcltx}
\usepackage{epsfig,graphicx,bm}
\usepackage{epstopdf, epsf}
\newcommand{\A}{\mathcal{A}}

\newcommand{\mpl}{M_{\rm Pl}}

\def\treh{T_{\rm reh}}
\newcommand{\mH}{\mathcal{H}}
\newcommand{\be}{\begin{equation}}
\newcommand{\ee}{\end{equation}}

\newcommand{\bse}{\begin{subequations}}
\newcommand{\ese}{\end{subequations}}
\newcommand{\bea}{\begin{eqnarray}}
\newcommand{\eea}{\end{eqnarray}}
\newcommand{\ba}{\begin{array}}
\newcommand{\ea}{\end{array}}
\newcommand{\bc}{\begin{center}}
\newcommand{\ec}{\end{center}}

\newcommand{\vev}[1]{\langle {#1}\rangle}
\newcommand{\N}{{\cal{N}}}

\definecolor{darkgreen}{rgb}{0,0.4,0}
\definecolor{darkblue}{rgb}{0,0,0.3}
\definecolor{darkred}{rgb}{0.7,0,0}

\definecolor{orange}{RGB}{255,100,0}

\begin{document}
\title{Leptogenesis in Inflationary models with Non-Abelian Gauge Fields}
\author{Azadeh Maleknejad}
\affiliation{School of Physics, Institute for Research in Fundamental Sciences (IPM),
P.~O.~Box 19395-5531,
Tehran, Iran}
\author{Mahdiyar Noorbala}
\affiliation{Department of Physics, University of Tehran,
North Karegar Ave., P.~O.~Box 14395-547, Tehran, Iran}
\author{M.~M.~Sheikh-Jabbari}
\affiliation{School of Physics, Institute for Research in Fundamental Sciences (IPM),
P.~O.~Box 19395-5531,
Tehran, Iran}
\preprint{IPM/P-2012/029}

\begin{abstract}

A scenario of leptogenesis was introduced in  \cite{Alexander:2004us} which works during inflationary period within standard model of particle physics setup. In this scenario lepton number is created by the gravitational chiral anomaly which has a non-zero expectation value for models of inflation driven by pseudoscalar field(s). Here, we observe that  models of inflation involving non-Abelian gauge fields, e.g.  the chromo-natural inflation \cite{Adshead:2012kp} or the gauge-flation \cite{gauge-flation-short}, have a parity-violating tensor mode (graviton) spectrum and naturally lead to a non-vanishing  expectation value for the gravitational chiral anomaly. Therefore, one has a natural leptogenesis scenario associated with these inflationary setups, \emph{inflato-natural leptogenesis}. We argue that the observed value of baryon-to-photon number density can be explained in a natural range of parameters in these models.

\end{abstract}

\maketitle

\section{Introduction}

Significant improvement in the cosmological observations and their precision in the last couple of decades has enabled us to partially uncover the
history of the early Universe. 
The leading paradigm is that the early Universe has undergone a period of accelerated expansion, {the} inflationary period,
followed by a reheating era leading to a radiation-dominated Universe and then matter-dominated period. Although we do not have the precise value of the
inflationary scale $H$ (Hubble during inflation) and the reheat temperature $\treh$, current observations provide an upper bound of $H\lesssim 10^{13}$ GeV, for the simplest single-field models, and the
strict lower bound of about $\treh\gtrsim 1$ MeV, due to the Big Bang Nucleosynthesis (BBN).  A particularly important prediction of the simplest inflationary models is the existence of primordial gravitational waves, which are not yet observed.  If observed, the amplitude of the primordial gravitational waves will fix the value of $H$ within these models.  On the other hand, if the primordial gravitational waves continue to evade detection, the upper bound on $H$ will improve, effectively ruling out more and more of the unsophisticated inflationary models.  We will stick to the inflationary paradigm for the rest of this paper.

A class of interesting questions one might ask is how details of {the} inflationary
model has affected the Universe we see today. Another class of interesting  questions is how sensitive inflationary models are to the UV, in particular
the Planck scale, physics. In this work we consider a particular model which has a feature relevant to  the first class of questions while this question is asked within two models of inflation which have a better controlled UV sensitivity behavior than most of the usual inflationary models. (For  possible implications of UV physics on CMB observations, e.g. see \cite{non-BD-Paper}.)

As far in the sky as we have observed, we seem not to have cosmological and astrophysical structures which are made out of antimatter; the observed Universe
consists of matter rather than antimatter. The matter-antimatter asymmetry given by the observations is usually quoted as \cite{Planck-mission}
\be\label{asymmetry}%
\eta=\frac{n_B-n_{\bar B}}{n_{\gamma}}\simeq 6 \times 10^{-10}\,,
\ee
where $n_B$, $n_{\bar B}$ and $n_\gamma$ are respectively the baryonic matter, antimatter and photon number densities in the observed Universe. Within the
inflationary setups the standard lore is that even if the matter-antimatter asymmetry is given by the initial conditions of the Universe, this asymmetry is
washed out by the rapid accelerated expansion of the Universe. Therefore, one should seek a \emph{dynamical} reasoning to explain the asymmetry.

About fifty years ago Sakharov \cite{Sakharov} formulated the three conditions needed for creating  matter-antimatter asymmetry from symmetric initial
conditions. Sakharov conditions demand existence of C and CP violation, baryon number violating interactions, and that these interactions should take place
out of equilibrium. Within the particle physics setups, it is generically easier to first create a lepton asymmetry (leptogenesis) and then relay on thermally
activated electroweak instantons (sphalerons) to create baryon asymmetry from the lepton asymmetry \cite{KRS85, KRS88, FY}. The sphalerons would be activated if the
temperature is not below $\sim 1-10$ TeV. Therefore, standard leptogenesis models demand a reheat temperature of around $\treh\gtrsim 10$ TeV, which in itself
and once we have a concrete reheating model, may impose a lower bound  on $H$.

Despite the fact that out-of-equilibrium condition is granted, inflationary period is not usually fit for leptogensis model building. This is because the
mechanisms used to provide the other two of Sakharov conditions are not generically efficient enough to compensate for the washout effect (exponential dilution) caused by
the rapid (usually almost exponential) expansion of the Universe during inflation. This obstacle, however, can be overcome if the mechanism for C, CP, and
lepton number violation is based on the fields which are active during inflation, i.e., metric and the inflaton(s). This is the idea put forward in
\cite{Alexander:2004us}, the gravi-leptogenesis.

The gravi-leptogensis is based on a particle physics model whose fermionic (chiral) matter content is assumed to be like that of  the Standard Model (SM),
with unequal number of left- and right- handed fermions. This model will have gravitational chiral anomaly \cite{AlvarezGaume:1983ig} on the $B-L$ current and hence there is room for
$B-L$ violation if
\be\label{RtildeR}
R\tilde R=\frac12\epsilon^{\alpha\beta\mu\nu}R_{\mu\nu\rho\sigma}R_{\alpha\beta}^{\ \ \ \rho\sigma}
\ee%
is nonzero on the background, or has a nonvanishing vacuum expectation value. This latter {case} can of course take place only if we have CP violation.
(Within the standard model C violation is already built in.) Therefore, all the three Sakharov conditions can be readily met within this setup. So, what is
left is to provide a setting in which $\langle R\tilde R\rangle\neq0$ can be realized during inflation.

It was noted in Ref.\cite{Alexander:2004us} that nonvanishing $\vev{R\tilde R}$ can be generated by tensor perturbations during inflation, if the spectrum of tensor modes is parity violating; i.e. if the two polarizations of tensor modes (gravitons) evolve differently. The setup  proposed in \cite{Alexander:2004us} to achieve this goal was to consider a pseudoscalar driven inflationary model (to provide the source for P and CP violation during inflation). The P and CP violation was then induced/transmitted  to the gravity sector through a coupling of the form $P(\chi) R\tilde R$, where $\chi$ is the pseudoscalar inflaton field and $P(\chi)$ is a generic odd function of $\chi$, which was added to the gravity action. In that setup   details of inflationary model was not relevant. Moreover,  it was argued that a $P(\chi)={\cal N}\frac{\chi}{\mpl}$ with ${\cal N}\sim 10^3$
naturally appears through supergravity or string theory compactifications involving axions \cite{Alexander:2007qe,St-Je}.

Here, we note that the P and CP violating inflationary background may be provided through inflationary models involving non-Abelian gauge fields; e.g. those discussed in \cite{gauge-flation-long,Gauge-flation-stability,AMW1,AMW2,Maleknejad:2012fw}. (This is of course a generic feature of inflationary models involving non-Abelian gauge fields \cite{Maleknejad:2014wsa}.) Interestingly, in these models usual minimal coupling of  non-Abelian gauge fields to gravity sector is enough for transmitting the P and CP violation in the inflationary sector to the tensor mode perturbations, and to have a nonvanishing
$\langle{R\tilde R}\rangle$. This observation suggests that i) inflationary models with non-Abelian gauge fields can be employed in building leptogenesis models, ii) one need not invoke $P(\chi) R\tilde R$-type interactions which requires large $P(\chi)$ values for a successful leptogenesis and hence one may relax the reliance on supergravity compactifications for that matter. (A similar idea, though with Abelian gauge fields and in a different setup, was analyzed in \cite{Stephon-David}.)

We would like to mention that the simplest setups of inflationary models with non-Abelain gauge fields have been disfavored by the Planck data \cite{Planck-mission}. Nonetheless, the main ingredient crucial to our discussion here, presence of intrinsic refrigerant gravitational waves, is a generic feature of inflationary models involving non-Abelian gauge fields \cite{Maleknejad:2014wsa}. It is hence still worthwhile to study
how the leptogenesis scenario outlined in \cite{Alexander:2004us} can be realized within non-Abelian gauge field driven models. In this work, however, for illustrative purposes we focus on the better studied gauge-flation and chromo-natural models.

This paper is organized as follows. We first very briefly review the chromo-natural model proposed in \cite{Adshead:2012kp} to provide the inflationary setup well-suited for realizing the inflato-natural leptogenesis mechanism outlined above. We  show how parity-violating tensor modes appear in this model and how they naturally lead to sizable $\langle{R\tilde R}\rangle$. We then compute lepton-antilepton asymmetry produced using gravitational anomaly. Relaying on sphalerons, this lepton-asymmetry is translated into Baryon-asymmetry, we can hence equate the computed lepton-number asymmetry to Baryon-asymmetry $n_B-n_{{\bar B}}$. Through a simple, but quite generic, reheating scenario we compute the photon number density $n_\gamma$ and finally discuss matching of this Baryon-asymmetry with the observed value \eqref{asymmetry}.

\section{Inflato-Natural Leptogenesis, the Setup}

As pointed out above, the inflato-natural leptogenesis consists of an inflationary sector which is taken to be the chromo-natural model (or gauge-flation) minimally coupled to Einstein gravity. The action describing this model is \cite{Adshead:2012kp}
\begin{equation}\label{The-action}
\begin{split}
{\cal L}&=-\frac{\mpl^2}{2}R-\frac{1}{4}F^a_{\mu\nu}F_a^{\mu\nu}-\frac12\partial_\mu\chi\partial^\mu\chi \\ &+\mu^4(1+\cos\frac{\chi}{f})
-\frac{\lambda}{8f}\chi F\tilde F,
\end{split}\end{equation}
where $\chi$ is the axion field, $F_{\mu\nu}$ is the field strength of an $SU(2)$ gauge theory and $F\tilde F =\epsilon^{\alpha\beta\mu\nu}F^a_{\mu\nu}F^a_{\alpha\beta}$. The spacetime indices will be denoted by Greek letters while the gauge indices by small Latin indices
$a,b,c=1,2,3$.  This model  has two
dimensionless parameters gauge coupling $g$ and axion-gauge field coupling $\lambda$, and two
dimensionful parameters $\mu$ and $f$.
Hereafter, we will work in units where $\mpl^{-2}=8\pi G=1$.

\subsection{Review of chromo-natural inflation}

To study  inflationary FLRW trajectories of chromo-natural model, we start with the following background metric and gauge field \cite{Adshead:2012kp} \begin{subequations}\begin{align}
ds^2 &=-dt^2+a^2(t) dx_idx_i\,,\\
\label{gauge-field}
A^a_\mu&=\left\{\begin{array}{cc} 0&\,,\qquad \mu=0,\\ a(t)\psi(t)\delta_i^a&\,,\qquad \mu=i .\end{array}\right.
\end{align}
\end{subequations}
With the above field configuration the rotation symmetry is retained, compensating the rotational  non-invariance caused by turning on vector gauge
fields in the background with the global part of the $SU(2)$ gauge symmetry group \cite{gauge-flation-short,gauge-flation-long,Galtsov91,Galtsov09,Galtsov10,Galtsov11};
$\psi(t)$ is a scalar under spatial rotations.

The slow-roll inflationary trajectories of the above model has been discussed in \cite{Maleknejad:2012fw,Mark-Peter,MaleknejadZarei}. For these trajectories
$\dot\chi/H\chi\sim \epsilon$, $\dot\psi/H\psi\sim \epsilon^2$, and during slow-roll inflation
\be\label{chipsi}
\begin{split}
\sin \frac{\chi}{f} \simeq& \frac{3g\lambda}{\mu^4}H\psi^3\,,\quad \epsilon\simeq \psi^2+\frac{3g^2\psi^4}{\mu^4(1+\cos\frac{\chi}{f})}\,, \\ 3H^2\simeq& \mu^4\left(1+\cos\frac{\chi}{f}\right)\,,
\quad \eta\simeq \psi^2\,,
\end{split}
\ee
where $H=\dot a/a$ is the Hubble parameter and, $\epsilon$ and $\eta$ are slow-roll parameters
\be
\epsilon=-\frac{\dot H}{H^2}\,,\qquad \eta=\frac{\ddot H}{2\dot H H}\,,
\ee
and $\simeq$ stands for equality up to the first order in $\epsilon,\ \eta$.

This model can lead to successful inflation for a wide range of its parameter space \cite{Mark-Peter,MaleknejadZarei}. Two regions which have been studied more thoroughly are
the small axion region $\chi_0/f\ll 1$  and large axion region $\chi_0/f\sim \pi$. (The large axion region the chromo-natural model is equivalent to gauge-flation model \cite{SheikhJabbari:2012qf}.)\footnote{Gauge-flation is two-parameter inflationary model described by the action \cite{gauge-flation-short}
\begin{equation}\label{gauge-flation-action}
{\cal L}=-\frac{\mpl^2}{2}R-\frac{1}{4}F^a_{\mu\nu}F_a^{\mu\nu}+\frac{\kappa}{384} (F\tilde F)^2\,.
\end{equation}
This model may be obtained from chromo-natural model upon integrating out the axion $\chi$ \cite{SheikhJabbari:2012qf}.}
Typical value of
parameters for these two regions are:

\noindent\textbf{Small axion model \cite{Mark-Peter,MaleknejadZarei}}
\begin{equation}
\label{chromo-natural}
\begin{aligned}
&\chi_0< 10^{-3}, & &f=10^{-2}, & &\lambda=200,\\
&H\simeq 8\times 10^{-8}, & &\mu^2=10^{-7}, & &g=2\times 10^{-6}, \\
&\psi \simeq 1.7\times 10^{-2}, & &\epsilon\simeq 10^{-3}, & &\eta\simeq 10^{-3}.
\end{aligned}
\end{equation}
\noindent\textbf{Large axion model \cite{SheikhJabbari:2012qf}}
\begin{equation}
\label{gauge-flation}
\begin{aligned}
&\chi_0-\pi f= 5\times 10^{-4}, & &f=10^{-2},& &\hspace*{-2mm} \lambda=2\times 10^4,\\
&H\simeq 3.3\times 10^{-5}, & &\hspace*{-8mm}\mu^2=1.6\times 10^{-3},& & g=10^{-3}, \\
&\psi\simeq 4\times 10^{-2}, & &\hspace*{-10mm}\epsilon\simeq 4\times 10^{-3},&  &\hspace*{-7mm}\eta\simeq 1.6\times 10^{-3}.
\end{aligned}
\end{equation}

\subsection{Tensor perturbations}

As discussed to run the inflato-natural leptogenesis machinery we need to analyze tensor perturbations. For the chromo-natural model this has been carried out in \cite{AMW1,AMW2}. There are two class of tensor perturbations in this model: the usual gravitons coming from metric perturbations,
\be
\delta g_{00}=\delta g_{0i}=0,\quad \textmd{and}\quad  \delta g_{ij}=a^2(t)h_{ij},
\ee
where $\delta g_{\mu\nu}$ denotes the perturbation of metric around FLRW metric, and
those coming from  the non-Abelian gauge field perturbations,
\be\label{pert-gf}
\delta A^a_0=0,\quad \textmd{and} \quad \delta A^a_i=a(t)\bigg(t_{ij}+\frac{\psi}{2} h_{ij}\bigg)\delta^{aj},
\ee
where $h_{ij}$ and $t_{ij}$ are symmetric, traceless and divergence-free tensor modes (for more details see  \cite{Maleknejad:2012fw}).

Using  perturbed Einstein equations, we obtain the field equation of $h_{ij}$, sourced by the contributions of $t_{ij}$ to the energy-momentum tensor
\be\label{GW-eq}
\Box h_{ij}=2\pi^T_{ij},
\ee
where $\Box$ is the d'Alamberian over the FLRW background and $\pi^T_{ij}$ is the tensor part of the anisotropic stress in the linear order energy-momentum tensor, \be \label{piT}
\pi^T_{~ij}\simeq2\psi\bigg(H^2(\gamma-1)t_{ij}-H\dot{t}_{ij}
+\sqrt{\gamma}H\partial_k t_{l(i}\epsilon^{~kl}_{j)}\bigg)\,.
\ee
Here $\simeq$ denotes equality at leading orders in slow-roll and $\gamma\equiv\frac{g^2\psi^2}{H^2}$.
Furthermore, the field equations of $t_{ij}$ is provided by the second order action of the tensor modes. Here, we summarize the noteworthy features of \eqref{GW-eq}, and the details may be found in \cite{Maleknejad:2012fw}:

\textbf{I.} {Compared to the usual scalar inflationary models in which $\pi^T_{ij}=0$, the field equation of $h_{ij}$ is modified by a nonvanishing anisotropic inertia.}

\textbf{II.} {The Chern-Simons interaction in the chromo-natural model (the last term in Eq.\eqref{The-action}) is a topological term which does not contribute to the energy-momentum tensor $T_{\mu\nu}$.\footnote{Likewise, the $(F\wedge F)^2$ term in the gauge-flation model \eqref{gauge-flation-action}, does not provide any contribution to the vector and tensor  perturbed energy-momentum tensor $\delta T_{\mu\nu}$.} As a result, $\pi^T_{ij}$ comes from the Yang-Mills term, as such, the
chromo-natural and gauge-flation models share the same tensor perturbations \cite{Maleknejad:2012fw}.

\textbf{III.} The last term in $\pi^T_{ij}$ is  parity odd and takes different signs for the two polarizations of the tensor modes, leading to the existence of intrinsic chiral gravitons in the setup \cite{AMW1,AMW2}. As a result, a nonvanishing $\vev{R\tilde R}$ is naturally generated within the chromo-natural and gauge-flation models.

To see the latter, we need to analyze the tensor perturbation equations. We will adopt the notations used in \cite{Maleknejad:2012fw}. The  divergence-free, traceless metric and gauge field perturbations can be conveniently parameterized in terms of the Left and Right polarization (helicity) modes,  $h_{_{R,L}},\ t_{_{R,L}}$. One may decompose the modes into Fourier modes of momentum $k$. For
our analysis below, as will become clear, we will only need the behavior of the modes in the deep inside horizon  ($k\!\gg\!aH$) region and here we will only focus on such modes. Field equations of the canonically normalized fields  $u_{_{R,L}}=\sqrt{2}~\!a~\!h_{_{R,L}}$ and $v_{_{R,L}}=2\sqrt{2}~\!a~\!t_{_{R,L}}$, are given as
\bea\
&&u''_{_{R,L}}+k^2 u_{_{R,L}}\simeq2\psi\mathcal{H}\bigg(-v'_{_{R,L}}\mp\sqrt{\gamma}k v_{_{R,L}}\bigg), \label{h-pert}\\
&&v''_{_{R,L}}+\bigg(k^2\mp2k\mathcal{H}\frac{(1+2\gamma)}{\sqrt{\gamma}}\bigg)v_{_{R,L}}\simeq0, \label{t-pert}
\eea
where prime denotes a derivative with respect to conformal time $\tau$, $d\tau=-dt/a(t)$ and $\mathcal{H}\equiv aH$.
Solving \eqref{h-pert} and \eqref{t-pert}, we obtain the following solution for $h_{_{R,L}}(k,\tau)$ in the deep inside horizon
\be\label{chiral-tensor-modes}
h_{_{R,L}}(k,\tau)\simeq-\frac{H\tau}{2\sqrt{k}}e^{-ik\tau}\bigg(1-\psi\big(\frac{\gamma\mp i\sqrt{\gamma}}{1+2\gamma}\big) e^{\mp i\frac{(1+2\gamma)}{\sqrt{\gamma}}\ln(-k\tau)}\bigg),
\ee
where we imposed usual Bunch-Davis normalization. Eq.\eqref{chiral-tensor-modes} indicates that the chromo-natural or gauge-flation models have the advocated intrinsic  birefringence at the level of gravity wave perturbations.
\vskip2mm

\subsection{Gravitational Anomaly, Causes Lepton and CP violation}

It is well-established in the literature \cite{AlvarezGaume:1983ig} that the gravitational chiral anomaly is
\begin{equation}\label{anomaly}
\nabla_\mu J^\mu_\ell  =  \frac{{\cal A}}{16\pi^2}  R  \tilde R
\end{equation}
where ${\cal A}=n_L-n_R$ measures the difference between number of left- and right-handed fermion degrees of freedom. For standard model ${\cal A}=3$,
while for beyond standard models with right-handed neutrinos it could be less than three.
To compute the total lepton number produced $L$, we follow the same lines as in \cite{Alexander:2004us}; i.e. Eq.\eqref{anomaly} should be viewed as an equality between two operators and we hence need to carry out quantization of the tensor modes. One can then readily deduce that
$$L=\frac{\A}{16\pi^2}\int_{-H^{-1}}^{\tau_f}d\tau \int d^3x \sqrt{-g} \vev{R\tilde R}\,,
$$
where $\tau_f$ is denotes end of inflation in the conformal time.

One may show that $\vev{R\tilde R}$, at second order in tensor perturbations $h_{_{L,R}}$ is basically the same as the classical expression for $R\tilde R$ evaluated for canonically normalized  tensor modes with Bunch-Davis initial state \cite{Maleknejad:2014wsa}. The details of this computation can be found in \cite{Maleknejad:2014wsa} and here we quote the final result
\[\begin{split}
n=\frac{\A}{8\pi^4 a^3(\tau_{_{f}})}\int^{\tau_{_{f}}}_{-H^{-1}} & d\tau \int_{\mH}^{k_{_{f}}(\tau)} k^3 dk \frac{d}{d\tau}\bigg(h'_R(\tau,k)h^{* '}_R(\tau,k)\cr &-k^2h_R(\tau,k)h^{*}_R(\tau,k)-R\leftrightarrow L\bigg)\,,
\end{split}\nonumber\]
where $n=L/(a^3\int d^3 x)$ is the lepton number density (per unit physical volume) and the integral over comoving momentum $k$ runs over all subhorizon \emph{quantum modes}, from the smallest physical momentum $H$  up to the UV cutoff momentum $k_{_{f}}(\tau)$: If we denote the cutoff on the physical momentum by $\Lambda$ and assume the slow-roll relation $a\simeq-1/(H\tau)$, then $k_{_{f}}(\tau)\simeq\Lambda/(H\tau)$. 
The restriction to subhorizon modes in the integral is due to the fact that we are calculating $L$ within a Hubble patch.  On the other hand, the presence of the UV cutoff is not only necessary to regulate the emerging infinity in the calculation, but also is required on the basis that our effective theory for quantum tensor fluctuations is only valid up to a finite energy scale, namely $\Lambda\ll\mpl$.
Inserting \eqref{chiral-tensor-modes} into the above integral we obtain
\be\label{Lepto-density}
n\simeq\frac{\A\mathcal{N}(\gamma)}{24\pi^4} \left(\frac{H}{\mpl}\right)^2 \frac{\psi}{\mpl} H^3 \left(\frac{\Lambda}{H}\right)^4\,.
\ee
The integral has been performed in $\Lambda\gg H$ limit and  $\mathcal{N}(\gamma)$ is an order one quantity; in the approximations we have used (e.g. dropping
the terms subleading in the subhorizon regime) it is given as
\[
\mathcal{N}(\gamma)\simeq \frac{(1+\gamma)(1+10\gamma)}{16\gamma+(1+2\gamma)^2}\bigg(
\frac{2\sqrt{\gamma}(2\gamma-1)}{1+10\gamma}
 \cos\beta-\sin\beta\bigg)
\,
\]
where $\beta=\frac{(1+2\gamma)\ln(\frac{\Lambda}{H})}{\sqrt{\gamma}}$.

\section{The $n/s$ ratio}

To compare with the observed data we need to compute the photon number density which up to a numerical factor is equal to the entropy density of the Universe \cite{KRS85, KRS88}. We do this with the standard assumption that the entropy of the Universe has not changed since the end of reheating. We also need a reheating model. Here, we  assume a slightly improved instant reheating model with a single ``refining'' or ``efficiency'' parameter $\sigma$:
\be\label{reheat}
\rho_{\rm reheat}=\sigma \rho_0=\frac{\pi^2}{30}g_* \treh^4\,,
\ee
where $g_*$ is the number of relativistic degrees of freedom and $\rho_0=3H^2\mpl^2$ is the energy density during inflation. $\sigma$ is a parameter which
measures the ``efficiency'' of the reheating process. For instant reheating, $\sigma=1$ (100\% efficiency) which leads to a typical reheat temperature
$\treh\sim 10^{14}$ GeV; or it can be as low as  $\sigma\sim 10^{-45}$ for $100$ TeV reheat temperatures. The entropy density $s$ is then
\be
s=\frac{2\pi^2}{45}g_* \treh^3= 2.3 g_*^{1/4}  \sigma^{3/4} (H\mpl)^{3/2}\,.
\ee

Finally, we can compute the desired $n/s$ {ratio}
\be\label{n/s}
\frac{n}{s}\simeq 9.7\times 10^{-4} \frac{\A\mathcal{N}(\gamma)}{g_{*}}\frac{\psi}{\mpl}\frac{H}{\mpl} \left(\frac{\mpl}{\treh}\right)^{3}\left(\frac{\Lambda}{\mpl}\right)^4.
\ee

Recalling the analysis of \cite{KRS85, KRS88} and the observations, the above should be compared with the observed value $n/s=8\times 10^{-11}$ \cite{Planck-mission}.

For typical values of $g_*\sim 10^2$, $\psi\sim 10^{-2}$,$\N\sim 1$ and setting $\A=3$, a successful leptogenesis model requires
\be
\left(\frac{\Lambda}{\mpl}\right)^4\left(\frac{\mpl}{\treh}\right)^3 \frac{H}{\mpl}\sim 10^{-3}.
\ee
This relation can be fulfilled for typically reasonable values of reheat temperature and cutoff $\Lambda$. For example, for $H\sim 10^{-5}\mpl,\ \Lambda\sim 10-100H$ and $\treh\sim 10^{12}-10^{13}$ GeV (consistent with \cite{Khlopov:1984pf}), we get a successful leptogenesis mechanism.  This corresponds to $\sigma\sim10^{-15}-10^{-11}$.

As we showed the ratio $n/s$ in our model crucially depends on the  reheating temperature. In our analysis we phenomenologically parametrized
the efficiency of the reheating model by the $\sigma$ parameter. It is desirable to study in more detail reheating within our gauge-flation and/or chromo-natural model.
In fact, as discussed in \cite{SheikhJabbari:2012qf} the two models become identical in the end of inflation.
Reheating in these models is also natural in
the sense that the energy of the system is already in the coherent oscillations of the gauge fields which could be taken to be gauge fields of standard model or beyond
and hence energy can directly be transferred to other standard model particles through gauge interactions.

\section*{Acknowledgement}

We would like to thank Stephon Alexander for useful comments. The final stage of this work was carried out during AM visit to the Max Planck Institute for Astrophysics and she greatly appreciates Eiichiro Komatsu for his warm hospitality and fruitful discussions.  M.N.\ acknowledges financial support from the research council of University of Tehran.


\section*{References}
\begin{thebibliography}{99}
\bibitem{Alexander:2004us}
S.~H.~-S.~Alexander, M.~E.~Peskin and M.~M.~Sheikh-Jabbari,
  Phys.\ Rev.\ Lett.\  {\bf 96}, 081301 (2006).

\bibitem{Adshead:2012kp}
  P.~Adshead, M.~Wyman,
 Phys.\ Rev.\ Lett.\  {\bf 108}, 261302 (2012).

\bibitem{gauge-flation-short}
  A.~Maleknejad, M.~M.~Sheikh-Jabbari,
 Phys.\ Lett.\ B {\bf 723}, 224 (2013)
  [arXiv:1102.1513 [hep-ph]].

\bibitem{non-BD-Paper}
  A.~Ashoorioon, K.~Dimopoulos, M.~M.~Sheikh-Jabbari and G.~Shiu,
arXiv:1306.4914 [hep-th].  

\bibitem{Planck-mission}
P.~A.~R.~Ade {\it et al.}  [Planck Collaboration],
  A\&A {\bf 594}, A13 (2016)
  [arXiv:1502.01589 [astro-ph.CO]].
  
\bibitem{Sakharov}
A.~D.~Sakharov,
Pisma Zh.\ Eksp.\ Teor.\ Fiz.\  {\bf 5}, 32 (1967)
[JETP Lett.\  {\bf 5}, 24 (1967)].


\bibitem{KRS85}
V.~A.~Kuzmin, V.~A.~Rubakov and M.~E.~Shaposhnikov,
Phys.\ Lett.\ B {\bf 155} (1985) 36.

\bibitem{KRS88}
S.~Y.~Khlebnikov and M.~E.~Shaposhnikov,
Nucl.\ Phys.\ B {\bf 308} (1988) 885.


\bibitem{FY}
M.~Fukugita, T.~Yanagida,
Phys.\ Lett.\ B {\bf 174} (1986) 45.

\bibitem{AlvarezGaume:1983ig}
  L.~Alvarez-Gaume and E.~Witten,
  Nucl.\ Phys.\ B {\bf 234}, 269 (1984).


\bibitem{Alexander:2007qe}
  S.~H.~-S.~Alexander, M.~E.~Peskin, M.~M.~Sheikh-Jabbari,
eConf C {\bf 0605151}, 0022 (2006)  [hep-ph/0701139].  

\bibitem{St-Je}
  S.~Alexander and J.~Martin,
  Phys.\ Rev.\ D {\bf 71}, 063526 (2005).

\bibitem{gauge-flation-long}
A. Maleknejad, M.M. Sheikh-Jabbari, 
Phys. Rev. D 84, 043515 (2011).

\bibitem{Gauge-flation-stability}
A.~Maleknejad, M.~M.~Sheikh-Jabbari and J.~Soda,
JCAP {\bf 1201} (2012) 016  [arXiv:1109.5573 [hep-th]].  

\bibitem{AMW1}
P.~Adshead, E.~Martinec and M.~Wyman,
 Phys.\ Rev.\ D {\bf 88}, 021302 (2013)
   [arXiv:1301.2598 [hep-th]]. 

\bibitem{AMW2}
P.~Adshead, E.~Martinec and M.~Wyman,
JHEP {\bf 1309} (2013) 087  [arXiv:1305.2930 [hep-th]].  


\bibitem{Maleknejad:2012fw}
  A.~Maleknejad, M.~M.~Sheikh-Jabbari and J.~Soda,
 Phys.\ Rept.\  {\bf 528}, 161 (2013)
  [arXiv:1212.2921 [hep-th]].

\bibitem{Maleknejad:2014wsa} 
  A.~Maleknejad,
  Phys.\ Rev.\ D {\bf 90}, 023542 (2014)
  [arXiv:1401.7628 [hep-th]].
  
\bibitem{Stephon-David}
  S.~Alexander, A.~Marciano and D.~Spergel,
 JCAP {\bf 1304}, 046 (2013)
  [arXiv:1107.0318 [hep-th]].


\bibitem{Galtsov91}
  D.~V.~Gal'tsov and M.~S.~Volkov,
Phys.\ Lett.\ B {\bf 256}, 17 (1991).
 
\bibitem{Galtsov09}
D.~V.~Gal'tsov,
arXiv:0901.0115 [gr-qc].  

\bibitem{Galtsov10}
D.~V.~Gal'tsov and E.~A.~Davydov,
Proc.\ Steklov Inst.\ Math.\  {\bf 272} (2011) 119  [arXiv:1012.2861 [gr-qc]].  

\bibitem{Galtsov11}
D.~V.~Gal'tsov and E.~A.~Davydov,
    arXiv:1112.2943 [hep-th].  


\bibitem{Mark-Peter}
P. Adshead and M. Wyman
  Phys.\ Rev.\ D {\bf 86}, 043530 (2012)
  [arXiv:1203.2264 [hep-th]].

\bibitem{MaleknejadZarei}
A. Maleknejad, M. Zarei,
  Phys.\ Rev.\ D {\bf 88}, 043509 (2013)
  [arXiv:1212.6760, arXiv:1212.6760].


\bibitem{SheikhJabbari:2012qf}
  M.~M.~Sheikh-Jabbari,
 Phys.\ Lett.\ B {\bf 717}, 6 (2012)
  [arXiv:1203.2265 [hep-th]].  

\bibitem{Khlopov:1984pf} 
  M.~Y.~.Khlopov and A.~D.~Linde,
  Phys.\ Lett.\ B {\bf 138}, 265 (1984).



\end{thebibliography}
\end{document}